\documentclass[11pt]{article}
\usepackage{graphicx}
\usepackage{amsmath}
\usepackage{amssymb}
\usepackage{caption2}
\begin{document}
\bibliographystyle{prsty}
\begin{center}
{\large {\bf \sc{  Reanalysis of  the heavy baryon states
$\Omega_b$, $\Omega_c$, $\Xi'_b$, $\Xi'_c$, $\Sigma_b$ and $\Sigma_c$ with QCD sum rules }}} \\[2mm]
Zhi-Gang Wang \footnote{E-mail,wangzgyiti@yahoo.com.cn.  }     \\
 Department of Physics, North China Electric Power University,
Baoding 071003, P. R. China

\end{center}

\begin{abstract}
In this article, we re-study the heavy baryon states $\Omega_b$,
$\Omega_c$, $\Xi'_b$, $\Xi'_c$, $\Sigma_b$ and $\Sigma_c$ with the
QCD sum rules, after subtracting the contributions from the
corresponding negative parity heavy baryon states, the predicted
masses  are in good agreement with the experimental data.
\end{abstract}



\section{Introduction}

The charm  and bottom baryons which  contain a heavy quark and two
light quarks are particularly interesting for studying dynamics of
the light quarks in the presence  of a heavy quark, and serve as an
excellent ground for testing predictions of the constituent quark
models and heavy quark symmetry. The ${1\over 2}^+$ antitriplet
states ($\Lambda_c^+$, $\Xi_c^+,\Xi_c^0)$,  and the ${1\over 2}^+$
and ${3\over 2}^+$ sextet states  ($\Omega_c,\Sigma_c,\Xi'_c$) and
($\Omega_c^*,\Sigma_c^*,\Xi^*_c$) have been established; while the
corresponding bottom  baryons are far from complete, only the
$\Lambda_b$, $\Sigma_b$, $\Sigma_b^*$, $\Xi_b$ and $\Omega_b$ have
been observed \cite{PDG}.

The QCD sum rules is a powerful theoretical tool in studying the
ground state heavy baryons \cite{SVZ79,PRT85}. The masses of the
$\Lambda_{Q}$, $\Sigma_{Q}$, $\Xi_{Q}$,
 $\Xi'_{Q}$, $\Omega_Q$,
 $\Sigma^*_{Q}$, $\Xi^*_{Q}$ and $\Omega^*_{Q}$
 have been calculated with the full QCD sum rules
 \cite{Bagan921,Bagan922,Nielsen07,Huang0805,Huang0811,Narison0904,Wang0809,Wang0910}.  The masses
of the $\Sigma^*_Q$, $\Sigma_Q$ and $\Lambda_Q$ have been calculated
with the QCD sum rules in the leading order of the heavy quark
effective theory  \cite{Shuryak82,Grozin92,Bagan93}, and later the
$1/m_Q$ corrections were  studied \cite{Dai961,Dai962,Huang02}.
Furthermore, the masses of the orbitally excited heavy baryons with
the leading order approximation \cite{Zhu00,HuangCS},
 and the $1/m_Q$ corrections \cite{HuangMQ} in the heavy
quark effective theory have also been analyzed. Recently the
$\frac{1}{2}^+$ and $\frac{3}{2}^+$  bottom  baryon states were
studied with the QCD sum rules in the heavy quark effective theory
including the $1/m_Q$ corrections \cite{Liu07}.

In 2008, the D0 collaboration reported the first observation of the
doubly strange baryon $\Omega_{b}^{-}$ in the decay channel
$\Omega_b^- \to J/\psi\thinspace\Omega^-$ (with
$J/\psi\to\mu^+\mu^-$ and $\Omega^-\to\Lambda K^-\to p\pi^- K^-$) in
$p\bar{p}$ collisions at $\sqrt{s}=1.96$ TeV \cite{OmegabD0}. The
experimental value $M_{\Omega_b^-}=6.165\pm 0.010\thinspace \pm
0.013\thinspace \, \rm{GeV}$ is about $0.1 \, \rm{GeV}$ larger than
the most theoretical calculations
\cite{Roncaglia95,Valcarce08,Jenkins96,Bowler96,Mathur02,Ebert05,Ebert08,Karliner07,Liu07,Roberts07}.
However,  the CDF collaboration did not confirm the   measured
 mass \cite{OmegabCDF}, i.e. they  observed the mass of the $\Omega^-_b$   is about $6.0544\pm 0.0068
\pm 0.0009 \,\rm{GeV} $, which is consistent with the most
theoretical calculations. On the other hand,  the theoretical
prediction $M_{\Omega_c^0}\approx 2.7\, \rm{GeV}$
\cite{Roncaglia95,Valcarce08,Jenkins96,Bowler96,Mathur02,Ebert05,Ebert08,Karliner07,Liu07,Roberts07,Liu0909}
is consistent with the experimental data
$M_{\Omega_c^0}=(2.6975\pm0.0026) \,\rm{GeV}$ \cite{PDG}.

In Ref.\cite{Oka96}, Jido et al introduce a novel approach based on
the QCD sum rules to separate the contributions of   the
negative-parity light flavor  baryons from the positive-parity light
flavor baryons, as the interpolating currents may have non-vanishing
couplings to both the negative- and positive-parity baryons
\cite{Chung82}. In Ref.\cite{Bagan93}, Bagan et al take the infinite
mass limit for the heavy quarks to  separate the contributions of
the positive and negative parity heavy baryon states  to the
correlation functions unambiguously before the work of Jido et al.
In this article, we re-study the masses and pole residues of the
${\frac{1}{2}}^+$ heavy baryon states $\Omega_Q$, $\Xi'_Q$ and
$\Sigma_Q$ by subtracting the contributions from the negative parity
baryon states. In Refs.\cite{Wang0704,Wang0809,Wang0910}, we study
the ${\frac{1}{2}}^+$ heavy baryons $\Omega_Q$, $\Xi'_Q$ and
$\Sigma_Q$ and ${\frac{3}{2}}^+$ heavy baryons $\Omega_Q^*$,
$\Xi^*_Q$ and $\Sigma^*_Q$ with the QCD sum rules in full QCD, and
observe that the pole residues of the ${\frac{3}{2}}^+$ heavy
baryons from the sum rules with different tensor structures are
consistent with each other, while the pole residues of the
${\frac{1}{2}}^+$ heavy baryons from the sum rules with different
tensor structures differ from each other greatly. Those pole
residues are important parameters in studying the radiative decays
$\Omega_Q^*\to \Omega_Q \gamma$, $\Xi_Q^*\to \Xi'_Q \gamma$ and
$\Sigma_Q^*\to \Sigma_Q \gamma$ \cite{Wang0910,Wang0909}, we should
refine those parameters to improve the predictive ability.

The article is arranged as follows:  we derive the QCD sum rules for
the masses and the pole residues of  the heavy baryon states
$\Omega_Q$, $\Xi'_Q$ and $\Sigma_Q$  in section 2; in section 3
numerical results are given and
 discussed,  and  section 4 is reserved for conclusion.

\section{QCD sum rules for  the $\Omega_Q$, $\Xi'_Q$ and $\Sigma_Q$}
The ${\frac{1}{2}}^+$ heavy baryons $\Omega_Q$, $\Xi'_Q$ and
$\Sigma_Q$ can be interpolated by the following currents
$J_\Omega(x)$, $J_\Xi(x)$ and $J_\Sigma(x)$ respectively,
\begin{eqnarray}
J_\Omega(x)&=& \epsilon^{ijk}  s^T_i(x)C\gamma_\mu s_j(x) \gamma_5 \gamma^\mu Q_k(x)  \, ,  \nonumber \\
J_\Xi(x)&=& \epsilon^{ijk}  q^T_i(x)C\gamma_\mu s_j(x) \gamma_5 \gamma^\mu Q_k(x)  \, ,  \nonumber \\
J_\Sigma(x)&=& \epsilon^{ijk}  u^T_i(x)C\gamma_\mu d_j(x) \gamma_5
\gamma^\mu Q_k(x)  \, ,
\end{eqnarray}
where the  $Q$ represents the heavy quarks $c$ and $b$,  the $i$,
$j$ and $k$ are color indexes, and the $C$ is the charge conjunction
matrix.  In this article, we take the simple Ioffe type
interpolating currents,  which are constructed by considering  the
diquark  theory  and the heavy quark symmetry
\cite{Jaffe2003,Jaffe2004}.

 The corresponding negative-parity heavy baryon states can be
interpolated by the  currents $J_{-} =i\gamma_{5} J_{+}$  because
multiplying $i \gamma_{5}$ to $J_{+}$ changes the parity of $J_{+}$
\cite{Oka96}, where the $J_{+}$ denotes the currents  $J_\Omega(x)$,
$J_\Xi(x)$ and $J_\Sigma(x)$. The correlation functions are defined
by
\begin{eqnarray}
\Pi_{\pm}(p)&=&i\int d^4x e^{ip \cdot x} \langle
0|T\left\{J_{\pm}(x)\bar{J}_{\pm}(0)\right\}|0\rangle \, ,
\end{eqnarray}
and can be decomposed as
\begin{equation}
    \Pi_{\pm}(p) = \!\not\!{p} \Pi_{1}(p^{2}) \pm \Pi_{0}(p^{2})\, ,
\end{equation}
due to Lorentz covariance,   because
\begin{equation}
    \Pi_{-}(p) = -\gamma_{5} \Pi_{+}(p)\gamma_{5}   \, .
\end{equation}
The currents $J_{+}$ couple  to both the positive-  and
negative-parity baryons \cite{Chung82},
\begin{eqnarray}
    \langle{0}|J_{+}| B^{-}\rangle \langle B^{-}|\bar{J}_{+}|0\rangle =
    - \gamma_{5}\langle 0|J_{-}| B^{-}\rangle \langle B^{-}| \bar{J}_{-}|0\rangle \gamma_{5} \, ,
\end{eqnarray}
where the $B^{-}$ denote the negative parity baryon states.

 We  insert  a
complete set  of intermediate baryon states with the same quantum
numbers as the current operators $J_{+}(x)$ and $J_{-}(x)$ into the
correlation functions $\Pi_{+}(p)$  to obtain the hadronic
representation \cite{SVZ79,PRT85}. After isolating the pole terms of
the lowest states, we obtain the following result \cite{Oka96}:
\begin{eqnarray}
    \Pi_{+}(p)     & = &   \lambda_+^2 {\!\not\!{p} +
    M_{+} \over M^{2}_+ -p^{2} } + \lambda_{-}^2
    {\!\not\!{p} - M_{-} \over M_{-}^{2}-p^{2}  } +\cdots \, ,
    \end{eqnarray}
where the $M_{\pm}$ are the masses of the lowest states with parity
$\pm$ respectively, and the $\lambda_{\pm}$ are the  corresponding
pole residues (or couplings).
 If we take $\vec{p} = 0$, then
\begin{eqnarray}
  \rm{limit}_{\epsilon\rightarrow0}\frac{{\rm Im}  \Pi_+(p_{0}+i\epsilon)}{\pi} & = &
    \lambda_+^2 {\gamma_{0} + 1\over 2} \delta(p_{0} - M_+) +
    \lambda_{-}^{2} {\gamma_{0} - 1\over 2} \delta(p_{0} - M_{-})+\cdots \nonumber \\
  & = & \gamma_{0} A(p_{0}) + B(p_{0})+\cdots \, ,
\end{eqnarray}
where
\begin{eqnarray}
  A(p_{0}) & = & {1 \over 2} \left[ \lambda_+^{2}
  \delta(p_{0} - M_+)  + \lambda_-^{2} \delta(p_{0} -
  M_{-})\right] \, , \nonumber \\
   B(p_{0}) & = & {1 \over 2} \left[ \lambda_+^{2}
  \delta(p_{0} - M_+)  - \lambda_-^{2} \delta(p_{0} -
  M_{-})\right] \, ,
\end{eqnarray}
the contribution $A(p_{0}) + B(p_{0})$ ($A(p_{0}) - B(p_{0})$)
contains contributions  from the positive parity (negative parity)
states only.

We carry out the operator product expansion at large $Q^2(=-p_0^2)$
region\footnote{We calculate the light quark parts of the
correlation functions $\Pi_{+}(p)$ in the coordinate space and use
the momentum space expression for the heavy quark propagators, then
resort to the Fourier integral to transform  the light quark parts
into the momentum space in $D$ dimensions, and take $\vec{p} = 0$.
For technical details, one can consult our previous works
\cite{Wang0704,Wang0809}.}, then use the dispersion relation to
obtain the spectral densities $\rho^A(p_0)$ and $\rho^B(p_0)$ (which
correspond to the tensor structures $\gamma_0$ and $1$ respectively)
at the level of quark-gluon degrees of freedom, finally we introduce
the weight functions $\exp\left[-\frac{p_0^2}{T^2}\right]$,
$p_0^2\exp\left[-\frac{p_0^2}{T^2}\right]$,   and obtain the
following sum rules,
\begin{eqnarray}
  \int_{\Delta}^{\sqrt{s_0}}dp_0 \left[
A(p_0)+B(p_0)\right]\exp\left[-\frac{p_0^2}{T^2}\right]&=&\int_{\Delta}^{\sqrt{s_0}}dp_0
\left[\rho^A(p_0)
+\rho^B(p_0)\right]\exp\left[-\frac{p_0^2}{T^2}\right] \,
,\nonumber \\
\end{eqnarray}
\begin{eqnarray}
  \int_{\Delta}^{\sqrt{s_0}}dp_0 \left[
A(p_0)+B(p_0)\right]
p_0^2\exp\left[-\frac{p_0^2}{T^2}\right]&=&\int_{\Delta}^{\sqrt{s_0}}dp_0
\left[\rho^A(p_0)
+\rho^B(p_0)\right]p_0^2\exp\left[-\frac{p_0^2}{T^2}\right] \,
,\nonumber \\
\end{eqnarray}
where the $s_0$ are the threshold parameters, $T^2$ is the Borel
parameter, $\Delta=m_Q+2m_s$, $\Delta=m_Q+m_s$ and $\Delta=m_Q$ in
the channels $\Omega_Q$, $\Xi'_Q$ and $\Sigma_Q$ respectively, the
explicit expressions of the spectral densities $\rho^A(p_0)$ and
$\rho^B(p_0)$ in the channels $\Omega_Q$, $\Xi'_Q$ and $\Sigma_Q$
are presented in the appendix. In calculation, we
 take  assumption of vacuum saturation for the high
dimension vacuum condensates, they  are always
 factorized to lower condensates with vacuum saturation in the QCD sum rules,
  and factorization works well in  large $N_c$ limit.
In this article, we take into account the contributions from the
quark condensates,  mixed condensates, gluon condensate, and neglect
the contributions  from other high dimension condensates, which are
suppressed by large denominators and would not play significant
roles.

\section{Numerical results and discussions}
The input parameters are taken to be the standard values $\langle
\bar{q}q \rangle=-(0.24\pm 0.01 \,\rm{GeV})^3$,  $\langle \bar{s}s
\rangle=(0.8\pm 0.2 )\langle \bar{q}q \rangle$, $\langle
\bar{q}g_s\sigma Gq \rangle=m_0^2\langle \bar{q}q \rangle$, $\langle
\bar{s}g_s\sigma Gs \rangle=m_0^2\langle \bar{s}s \rangle$,
$m_0^2=(0.8 \pm 0.2)\,\rm{GeV}^2$ \cite{Ioffe2005,LCSRreview},
$\langle \frac{\alpha_s GG}{\pi}\rangle=(0.012 \pm
0.004)\,\rm{GeV}^4 $ \cite{LCSRreview},
$m_s=(0.14\pm0.01)\,\rm{GeV}$, $m_c=(1.35\pm0.10)\,\rm{GeV}$ and
$m_b=(4.7\pm0.1)\,\rm{GeV}$ \cite{PDG} at the energy scale  $\mu=1\,
\rm{GeV}$.

Those vacuum condensates can be  calculated with
 lattice QCD and  instanton  models, or determined by fitting certain
QCD sum rules to the experimental data; the values are consistent
with each other (except for the gluon condensate $\langle
\frac{\alpha_s GG}{\pi}\rangle $) considering the uncertainties. The
value of the gluon condensate $\langle \frac{\alpha_s
GG}{\pi}\rangle $ has been updated from time to time, and changes
greatly (for a comprehensive review, one can consult the book "QCD
as a theory of hadrons from partons to confinement" by S.Narison
\cite{NarisonBook}).
 At the present case, the gluon condensate  makes tiny  contribution,  the updated value $\langle \frac{\alpha_s GG}{\pi}\rangle=(0.023 \pm
0.003)\,\rm{GeV}^4 $ \cite{NarisonBook} and the standard value
$\langle \frac{\alpha_s GG}{\pi}\rangle=(0.012 \pm
0.004)\,\rm{GeV}^4 $ \cite{LCSRreview} lead to a difference less
than $2\,\rm{MeV}$ for the masses.

The $Q$-quark masses appearing in the perturbative terms (see the
appendix) are usually taken to be the pole masses in the QCD sum
rules, while the choice of the $m_Q$ in the leading-order
coefficients of the higher-dimensional terms is arbitrary
\cite{NarisonBook,Kho9801}. For example, the $\overline{MS}$ mass
$m_c(m_c^2)$ relates with the pole mass $\hat{m}$ through the
relation
\begin{eqnarray}
m_c(m_c^2) &=&\hat{m}\left[1+\frac{C_F
\alpha_s(m_c^2)}{\pi}+(K-2C_F)\left(\frac{\alpha_s}{\pi}\right)^2+\cdots\right]^{-1}\,
,
\end{eqnarray}
where $K$ depends on the flavor number $n_f$. In this article, we
take the approximation $m_c\approx\hat{m}$ without the $\alpha_s$
corrections for consistency. The value listed in the Particle Data
Group is $m_c(m_c^2)=1.27^{+0.07}_{-0.11} \, \rm{GeV}$ \cite{PDG},
it is reasonable to take the value
$m_c=m_c(1\,\rm{GeV}^2)=(1.35\pm0.10)\,\rm{GeV}$ in our works. The
mass of the $b$ quark $m_b$ can be understood   analogously.

In calculation, we  also neglect  the contributions from the
perturbative corrections $\mathcal {O}(\alpha_s^n)$.  Those
perturbative corrections can be taken into account in the leading
logarithmic
 approximations through  anomalous dimension factors. After the Borel transform, the effects of those
 corrections are  to multiply each term on the operator product
 expansion side by the factor,
 \begin{eqnarray}
 \left[ \frac{\alpha_s(T^2)}{\alpha_s(\mu^2)}\right]^{2\Gamma_{J}-\Gamma_{\mathcal
 {O}_n}} \, ,
 \end{eqnarray}
 where the $\Gamma_{J}$ is the anomalous dimension of the
 interpolating current $J(x)$, the $\Gamma_{\mathcal {O}_n}$ is the anomalous dimension of
 the local operator $\mathcal {O}_n(0)$ in the operator product
 expansion,
 \begin{eqnarray}
 T\left\{J(x)J^{\dagger}(0)\right\}&=&C_n(x) {O}_n(0) \, ,
 \end{eqnarray}
here the $C_n(x)$ is the corresponding Wilson coefficient.

We carry out the operator product expansion at a special energy
scale $\mu^2=1\,\rm{GeV}^2$, and  set the factor $\left[
\frac{\alpha_s(T^2)}{\alpha_s(\mu^2)}\right]^{2\Gamma_{J}-\Gamma_{\mathcal
{O}_n}}\approx1$, such an approximation maybe result in some scale
dependence  and  weaken the prediction ability. In this article, we
study the $J^P=\frac{1}{2}^+$ sextet heavy baryon states
($\Omega_c,\Sigma_c,\Xi'_c$) and ($\Omega_b,\Sigma_b,\Xi'_b$)
systemically, and can reproduce the masses of the  well established
baryon states, the predictions are still robust as we take the
analogous criteria in those sum rules.

In the conventional QCD sum rules \cite{SVZ79,PRT85}, there are two
criteria (pole dominance and convergence of the operator product
expansion) for choosing  the Borel parameter $T^2$ and threshold
parameter $s_0$.  We impose the two criteria on the heavy baryon
states to choose the Borel parameter $T^2$ and threshold parameter
$s_0$, the values are shown in Table 1. From Table 1, we can see
that the contribution from the perturbative term  is dominant, the
operator product expansion is convergent certainly. In this article,
we take the contribution from the pole term is larger than $45\%$,
the uncertainty of the threshold parameter is  $0.1\,\rm{GeV}$, and
the Borel window is $1\,\rm{GeV}^2$.

Taking into account all uncertainties  of the  parameters,  we
obtain the values of the masses and pole residues of
 the  heavy baryon states, which are  shown in Figs.1-2 and Table 2.

\begin{table}
\begin{center}
\begin{tabular}{|c|c|c|c|c|c|}
\hline\hline & $T^2 (\rm{GeV}^2)$& $\sqrt{s_0} (\rm{GeV})$&pole&perturbative\\
\hline
 $\Omega_b$  &$5.2-6.2$ &$6.8$&  $(45-60)\%$&$(83-88)\%$\\ \hline
       $\Xi'_b$  &$4.9-5.9$ &$6.7$& $(45-61)\%$&$(77-84)\%$\\ \hline
           $\Sigma_b$  &$4.6-5.6$ & $6.6$& $(45-63)\%$&$(70-81)\%$\\ \hline
           $\Omega_c$  &$2.2-3.2$ &$3.4$& $(46-75)\%$ &$(75-86)\%$\\ \hline
              $\Xi'_c$  &$2.0-3.0$ &$3.3$ & $(47-78)\%$&$(67-83)\%$\\ \hline
              $\Sigma_c$  &$1.8-2.8$ &$3.2$& $(47-82)\%$&$(57-79)\%$\\ \hline
    \hline
\end{tabular}
\end{center}
\caption{ The Borel parameters $T^2$ and threshold parameters $s_0$
for the heavy baryon states, the "pole" stands for the contribution
from the pole term, and the "perturbative" stands for the
contribution from the perturbative term in the operator product
expansion.}
\end{table}

\begin{table}
\begin{center}
\begin{tabular}{|c|c|c|c|c|c|c|}
\hline\hline & $T^2 (\rm{GeV}^2)$& $\sqrt{s_0} (\rm{GeV})$&
$M(\rm{GeV})$&$\lambda (\rm{GeV}^3)$&$M(\rm{GeV})[\rm{exp}]$\\\hline
 $\Omega_b$  &$5.2-6.2$ &$6.8\pm0.1$&  $6.11\pm0.16$&$0.134\pm0.030$&$6.165\cite{OmegabD0}/6.0544\cite{OmegabCDF}$\\ \hline
       $\Xi'_b$  &$4.9-5.9$ &$6.7\pm0.1$& $5.96\pm0.17$&$0.079\pm0.020$&?\\ \hline
           $\Sigma_b$  &$4.6-5.6$ & $6.6\pm0.1$& $5.80\pm0.19$&$0.062\pm0.018$&5.8078($\Sigma_b^+$)/5.8152($\Sigma_b^-$) \cite{PDG}\\ \hline
           $\Omega_c$  &$2.2-3.2$ &$3.4\pm0.1$& $2.70\pm0.20$&$0.093\pm0.023$ &$2.6952$\cite{PDG}\\ \hline
              $\Xi'_c$  &$2.0-3.0$ &$3.3\pm0.1$ & $2.56\pm0.22$&$0.055\pm0.016$&2.5756($\Xi^{'+}_c$)/2.5779($\Xi^{'0}_c$)\cite{PDG}\\ \hline
              $\Sigma_c$  &$1.8-2.8$ &$3.2\pm0.1$& $2.40\pm0.26$&$0.045\pm0.015$&2.454\cite{PDG}\\ \hline
    \hline
\end{tabular}
\end{center}
\caption{ The masses $M(\rm{GeV})$ and pole residues
$\lambda(\rm{GeV}^3)$ of the heavy baryon states.}
\end{table}

\begin{figure}
 \centering
 \includegraphics[totalheight=5cm,width=6cm]{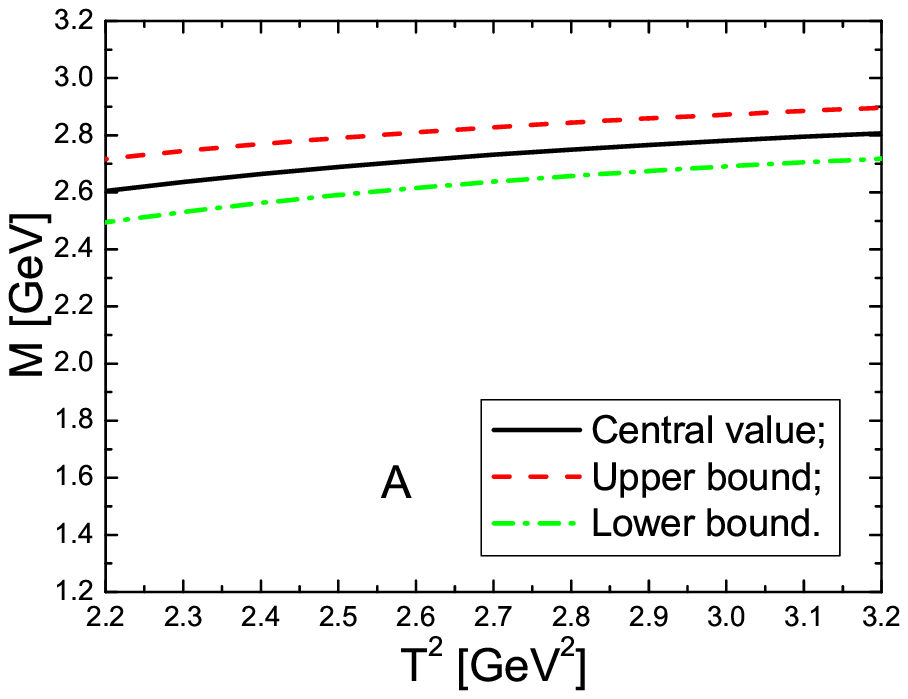}
 \includegraphics[totalheight=5cm,width=6cm]{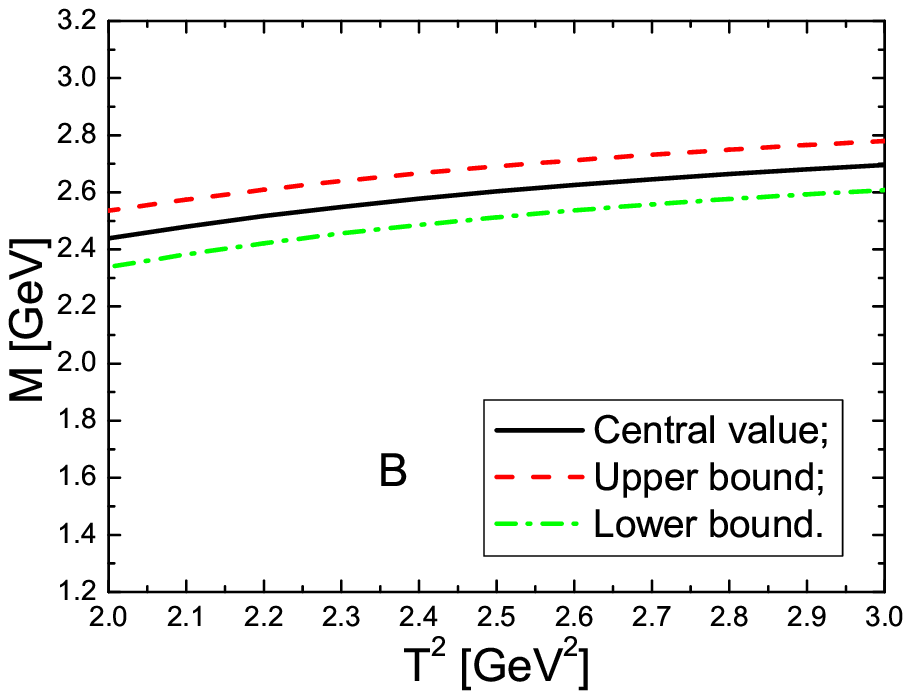}
 \includegraphics[totalheight=5cm,width=6cm]{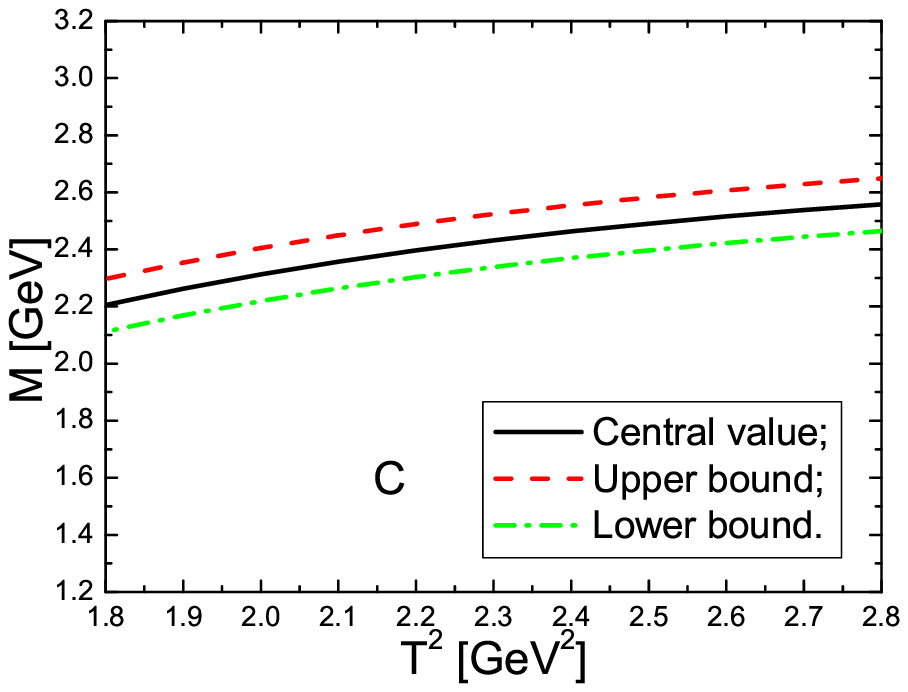}
 \includegraphics[totalheight=5cm,width=6cm]{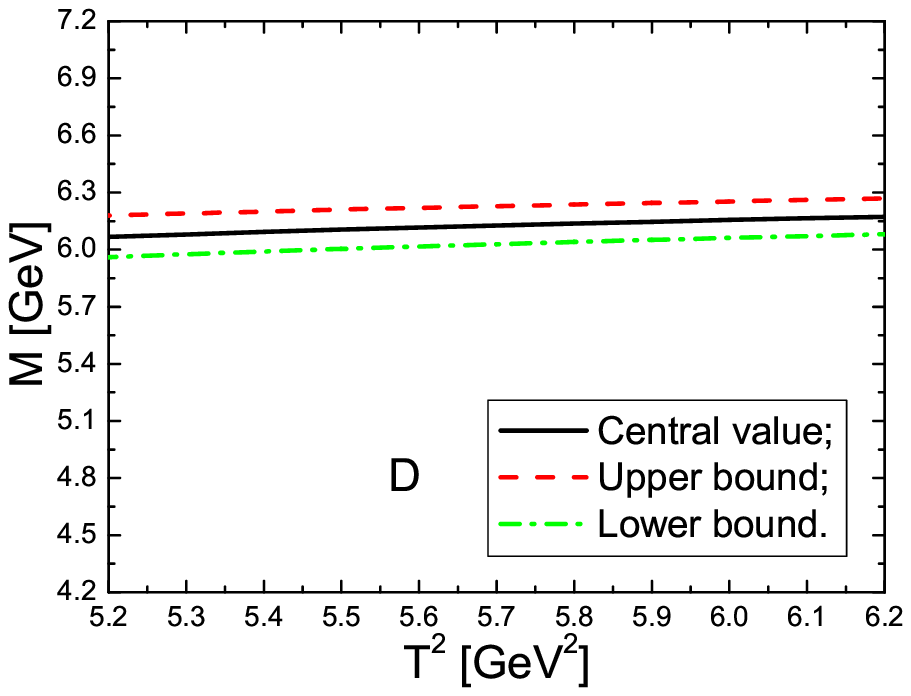}
 \includegraphics[totalheight=5cm,width=6cm]{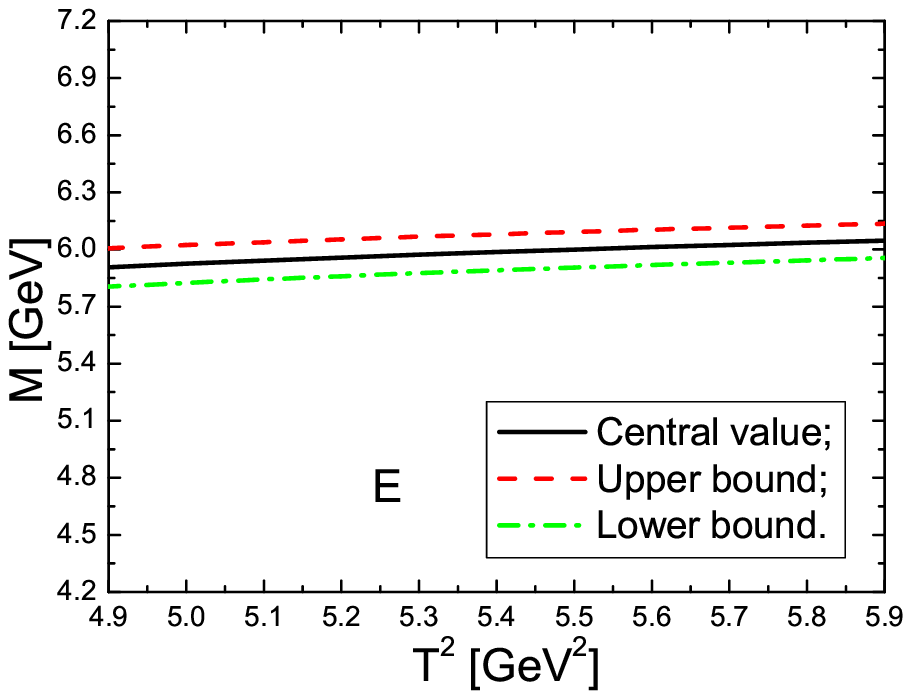}
 \includegraphics[totalheight=5cm,width=6cm]{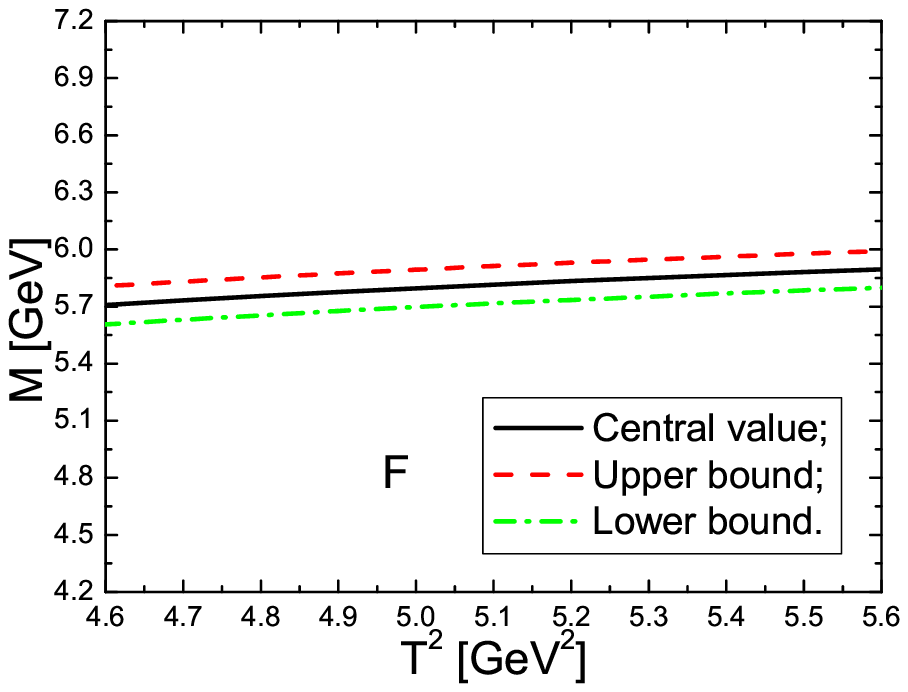}
       \caption{ The masses  $M$ of the heavy baryon states, the $A$, $B$, $C$, $D$, $E$ and $F$ correspond
       to the channels $\Omega_c$, $\Xi'_c$, $\Sigma_c$, $\Omega_b$, $\Xi'_b$ and $\Sigma_b$ respectively.  }
\end{figure}

\begin{figure}
 \centering
 \includegraphics[totalheight=5cm,width=6cm]{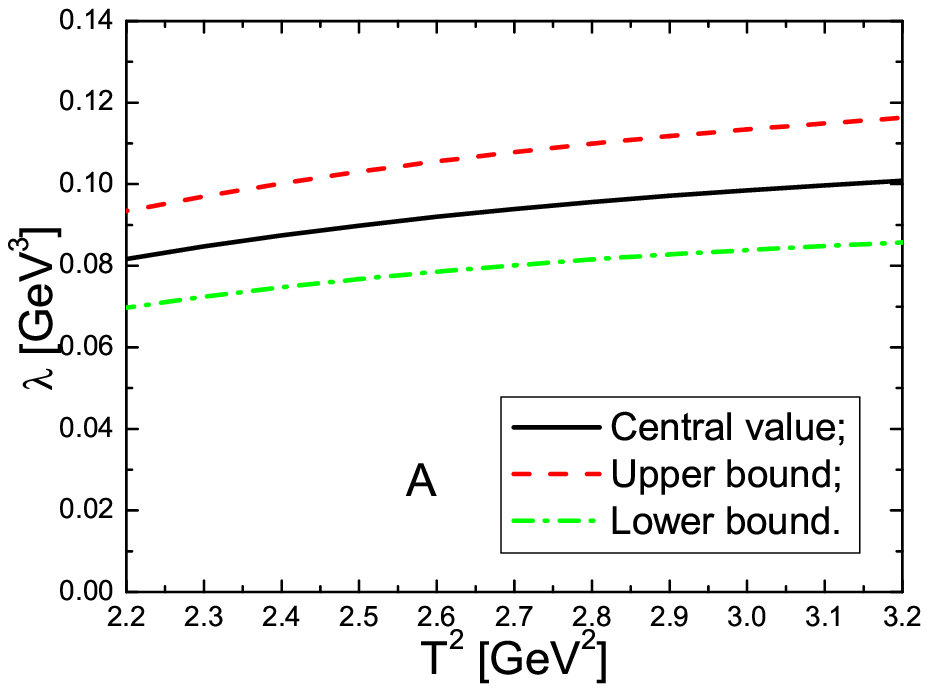}
  \includegraphics[totalheight=5cm,width=6cm]{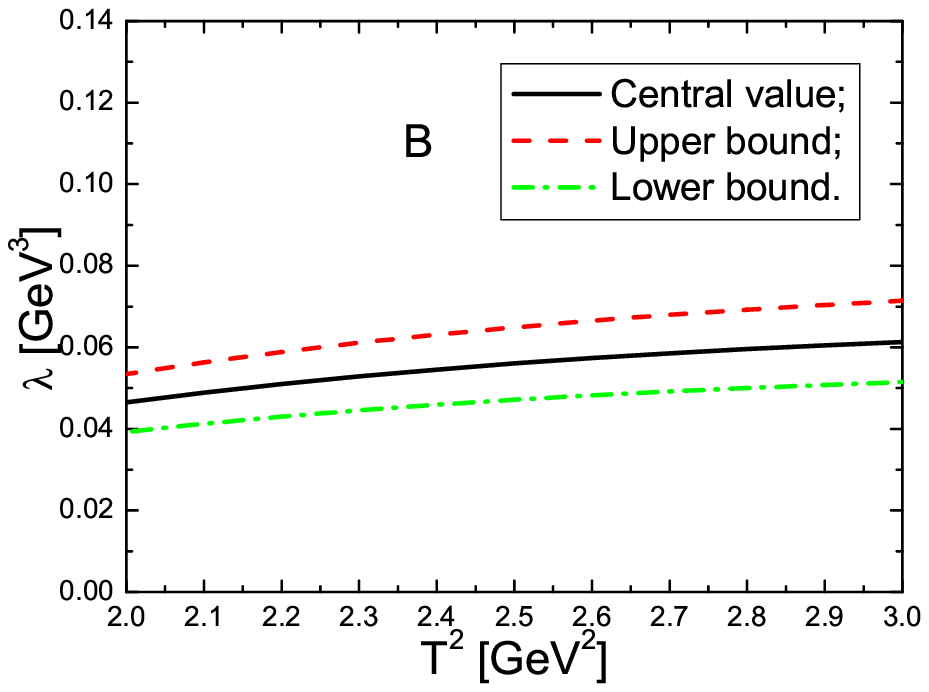}
    \includegraphics[totalheight=5cm,width=6cm]{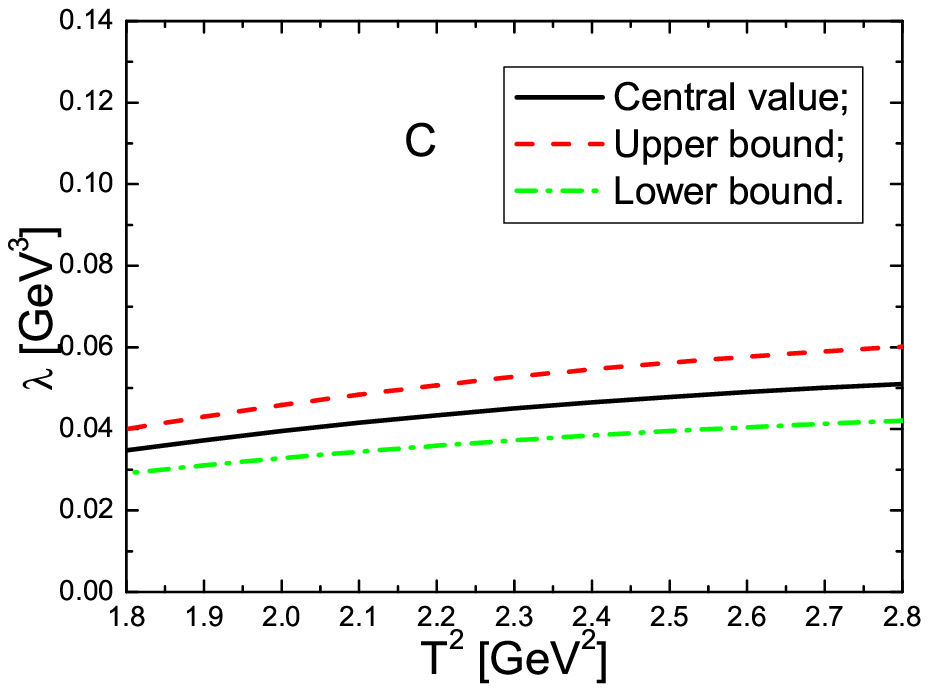}
     \includegraphics[totalheight=5cm,width=6cm]{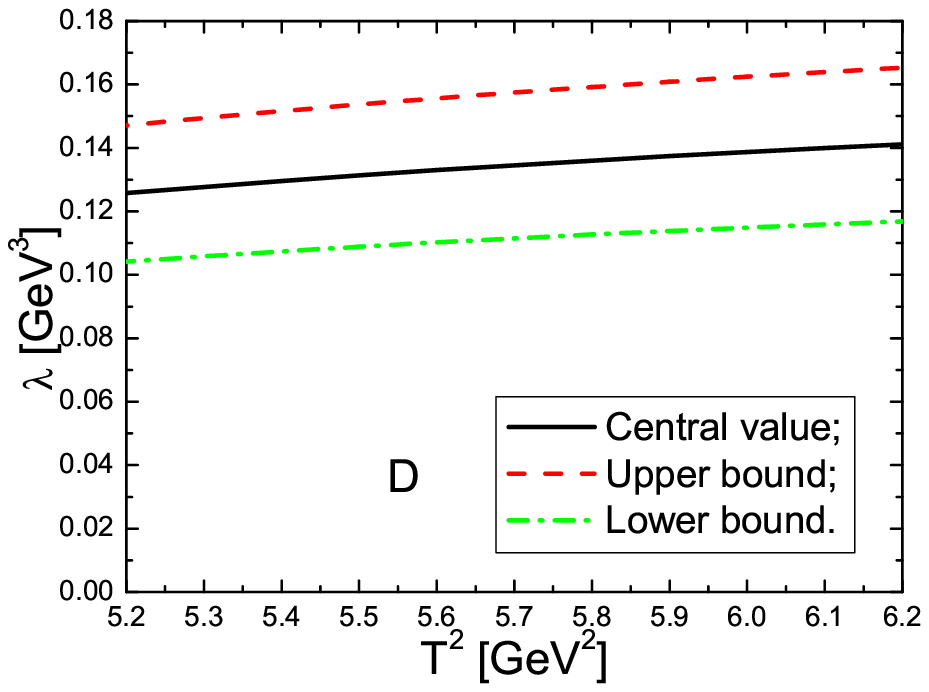}
      \includegraphics[totalheight=5cm,width=6cm]{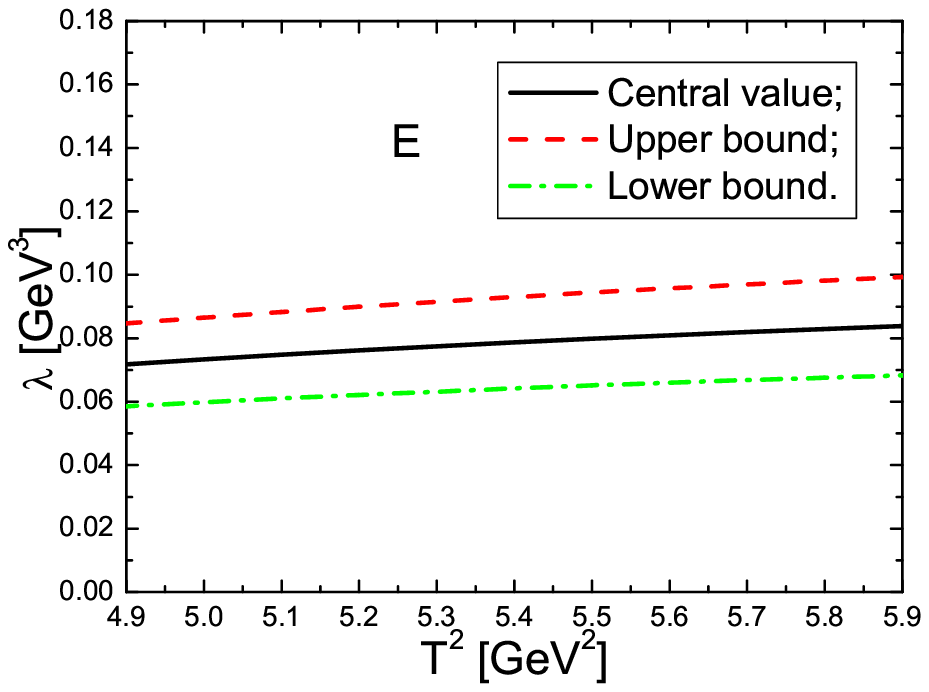}
       \includegraphics[totalheight=5cm,width=6cm]{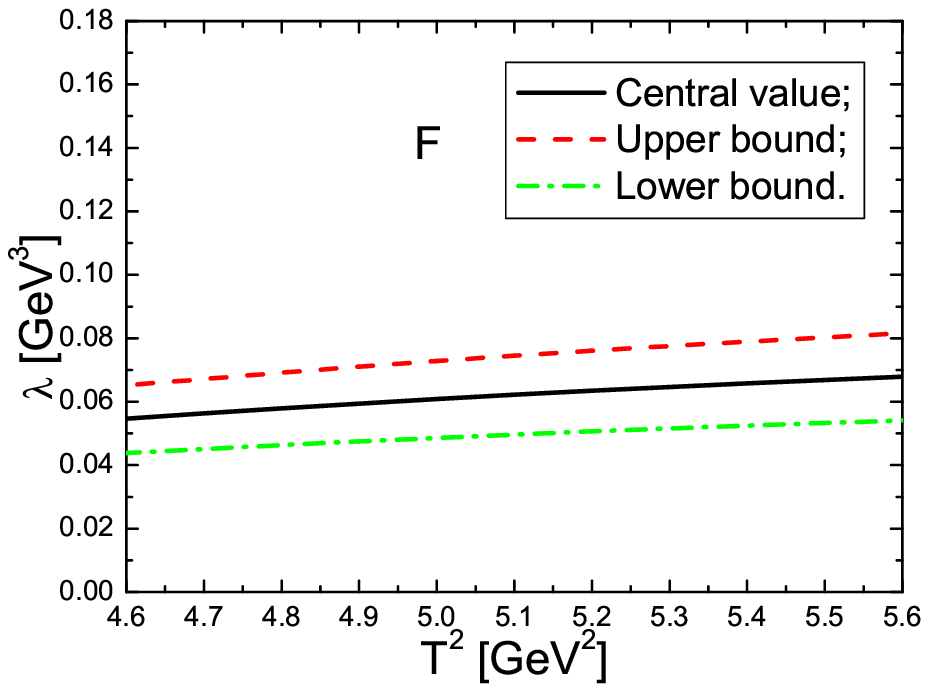}
       \caption{ The pole residues $\lambda$ of the heavy baryon states, the $A$, $B$, $C$, $D$, $E$ and $F$ correspond
       to the channels $\Omega_c$, $\Xi'_c$, $\Sigma_c$, $\Omega_b$, $\Xi'_b$ and $\Sigma_b$ respectively.  }
\end{figure}

\begin{figure}
 \centering
 \includegraphics[totalheight=5cm,width=6cm]{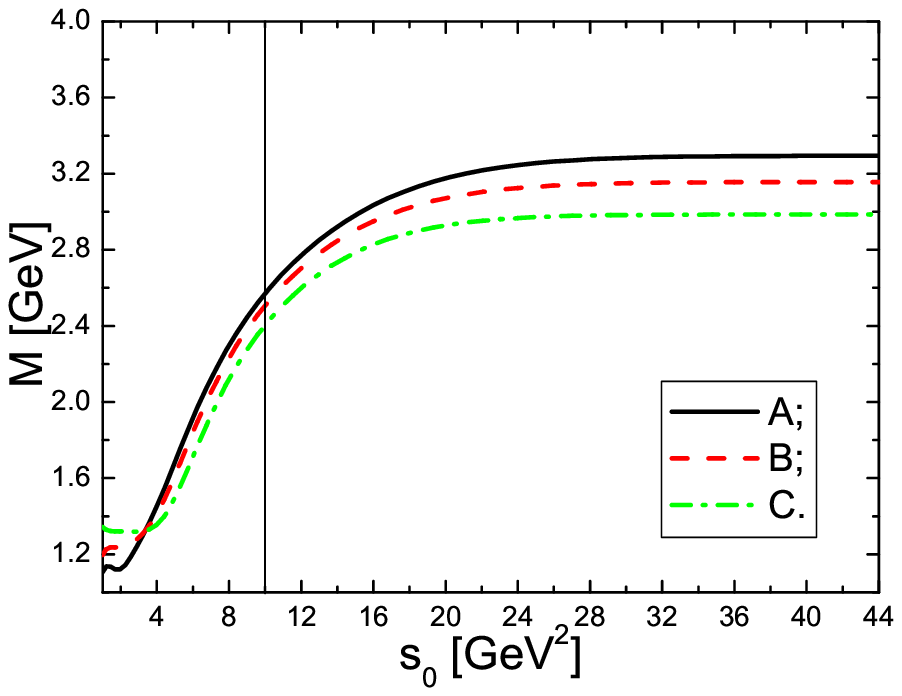}
        \includegraphics[totalheight=5cm,width=6cm]{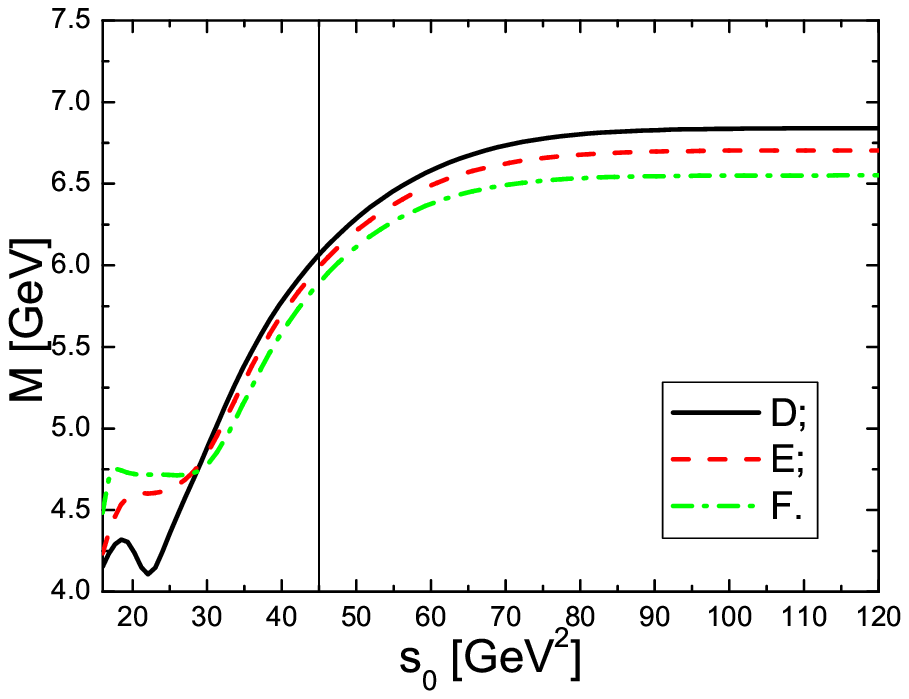}
       \caption{ The predicted masses with variation of the threshold parameters $s_0$, where the Borel parameters are taken to be  the central values presented in Table 2, the $A$, $B$, $C$, $D$, $E$ and $F$ correspond
       to the channels $\Omega_c$, $\Xi'_c$, $\Sigma_c$, $\Omega_b$, $\Xi'_b$ and $\Sigma_b$ respectively.  }
\end{figure}

From Table 2, we can see that the present predictions for the masses
of the heavy baryons are consistent with the experimental data, the
energy gap among  the central values of the present predictions is
about $M_{\Omega_b}-M_{\Xi'_b}\approx M_{\Xi'_b}-M_{\Sigma_b}\approx
M_{\Omega_c}-M_{\Xi'_c}\approx M_{\Xi'_c}-M_{\Sigma_c}\approx
0.15\,\rm{GeV}$, which is excellent. The central value
$M_{\Omega_b}=6.11\,\rm{GeV}$ lies between the experimental data
$6.165\,\rm{GeV}$ from the D0 collaboration \cite{OmegabD0} and
$6.0544\,\rm{GeV}$ from the CDF collaboration \cite{OmegabCDF}. More
experimental data is still needed to confirm the present
predictions.

In Fig.3, we plot the  predicted masses with variation of the
threshold parameters $s_0$. From the figure, we can see that the
predicted masses increase almost linearly with the threshold
parameters $s_0$  for $s_0\leq 10\,\rm{GeV}^2$ in the charm channels
and $s_0\leq 45\,\rm{GeV}^2$ in the bottom channels respectively.
The threshold parameters should be taken around the
 critical points which are shown by the vertical lines  in Fig.3, the values we
 choose in Table 2 are reasonable.

\section{Conclusion}
In this article, we re-study the heavy baryon states $\Omega_Q$,
$\Xi'_Q$ and $\Sigma_Q$ with the QCD sum rules, after subtracting
the contributions from the corresponding negative parity heavy
baryon sates, the predicted masses are in good agreement with the
experimental data.

\section*{Appendix}
 The spectral densities of the heavy baryon
states $\Omega_Q$, $\Xi'_Q$ and $\Sigma_Q$ at the level of
quark-gluon degrees of freedom,

\begin{eqnarray}
\rho^A_{\Omega_Q}(p_0)&=&\frac{p_0}{16\pi^4}\int_{t_i}^1dt
t(1-t)^3(p_0^2-\widetilde{m}_Q^2)(5p_0^2-3\widetilde{m}_Q^2)-\frac{p_0m_s\langle\bar{s}s\rangle}{\pi^2}\int_{t_i}^1
dt t\nonumber\\
&&+\frac{m_s\langle\bar{s}s\rangle}{\pi^2}\int_{t_i}^1 dt
t(1-t)\left[3p_0+\frac{\widetilde{m}_Q^2}{2}\delta
(p_0-\widetilde{m}_Q)\right]
\nonumber\\
&&-\frac{m_s\langle\bar{s}g_s\sigma Gs\rangle}{6\pi^2}\int_0^1dt
t\left[1+\frac{\widetilde{m}_Q}{4T}\right]\delta
(p_0-\widetilde{m}_Q) \nonumber\\
&&+\frac{m_s\langle\bar{s}g_s\sigma Gs\rangle}{8\pi^2}\delta
(p_0-m_Q)+\frac{\langle\bar{s}s\rangle^2}{3}\delta(p_0-m_Q)\nonumber \\
&& +\frac{p_0}{48\pi^2}\langle \frac{\alpha_sGG}{\pi}\rangle
\int_{t_i}^1 dt (4-5t)\nonumber \\
&&+\frac{1}{96\pi^2}\langle \frac{\alpha_sGG}{\pi}\rangle
\int_{t_i}^1 dt (1-t)\widetilde{m}_Q^2\delta
(p_0-\widetilde{m}_Q)\nonumber \\
&& -\frac{m_Q^2}{144\pi^2}\langle \frac{\alpha_sGG}{\pi}\rangle
\int_0^1 dt\frac{(1-t)^3}{t^2}\left[1+\frac{p_0}{4T}\right]\delta
(p_0-\widetilde{m}_Q) \, ,
\end{eqnarray}

\begin{eqnarray}
\rho^B_{\Omega_Q}(p_0)&=&\frac{3m_Q}{32\pi^4}\int_{t_i}^1dt
(1-t)^2(p_0^2-\widetilde{m}_Q^2)^2-\frac{3m_sm_Q\langle\bar{s}s\rangle}{2\pi^2}\int_{t_i}^1
dt
\nonumber\\
&&+\frac{5m_s \langle\bar{s}g_s\sigma Gs\rangle }{24\pi^2}\delta
(p_0-m_Q) +\frac{2\langle\bar{s}s\rangle^2}{3}\delta(p_0-m_Q)\nonumber\\
&&+\frac{m_Q}{96\pi^2}\langle \frac{\alpha_sGG}{\pi}\rangle
\int_{t_i}^1 dt \left[ -3-2t+\frac{2}{t^2}\right]\nonumber \\
&& -\frac{m_Q}{192\pi^2}\langle \frac{\alpha_sGG}{\pi}\rangle
\int_0^1 dt\frac{(1-t)^2}{t}\widetilde{m}_Q \delta
(p_0-\widetilde{m}_Q) \, ,
\end{eqnarray}

\begin{eqnarray}
\rho^A_{\Xi'_Q}(p_0)&=&\frac{p_0}{32\pi^4}\int_{t_i}^1dt
t(1-t)^3(p_0^2-\widetilde{m}_Q^2)(5p_0^2-3\widetilde{m}_Q^2)-\frac{p_0m_s\langle\bar{q}q\rangle}{4\pi^2}\int_{t_i}^1
dt t\nonumber\\
&&+\frac{m_s\langle\bar{s}s\rangle}{4\pi^2}\int_{t_i}^1 dt
t(1-t)\left[3p_0+\frac{\widetilde{m}_Q^2}{2}\delta
(p_0-\widetilde{m}_Q)\right]
\nonumber\\
&&-\frac{m_s\langle\bar{s}g_s\sigma Gs\rangle}{24\pi^2}\int_0^1dt
t\left[1+\frac{\widetilde{m}_Q}{4T}\right]\delta
(p_0-\widetilde{m}_Q ) \nonumber\\
&&+\frac{m_s\langle\bar{q}g_s\sigma Gq\rangle}{32\pi^2}\delta
(p_0-m_Q)+\frac{\langle\bar{q}q\rangle\langle\bar{s}s\rangle}{6}\delta(p_0-m_Q)\nonumber \\
&& +\frac{p_0}{96\pi^2}\langle \frac{\alpha_sGG}{\pi}\rangle
\int_{t_i}^1 dt (4-5t)\nonumber \\
&&+\frac{1}{192\pi^2}\langle \frac{\alpha_sGG}{\pi}\rangle
\int_{t_i}^1 dt (1-t)\widetilde{m}_Q^2\delta
(p_0-\widetilde{m}_Q)\nonumber \\
&& -\frac{m_Q^2}{288\pi^2}\langle \frac{\alpha_sGG}{\pi}\rangle
\int_0^1
dt\frac{(1-t)^3}{t^2}\left[1+\frac{\widetilde{m}_Q}{4T}\right]\delta
(p_0-\widetilde{m}_Q) \, ,
\end{eqnarray}

\begin{eqnarray}
\rho^B_{\Xi'_Q}(p_0)&=&\frac{3m_Q}{64\pi^4}\int_{t_i}^1dt
(1-t)^2(p_0^2-\widetilde{m}_Q^2)^2-\frac{m_sm_Q\langle\bar{q}q\rangle}{2\pi^2}\int_{t_i}^1
dt+\frac{m_sm_Q\langle\bar{s}s\rangle}{8\pi^2}\int_{t_i}^1 dt
\nonumber\\
&&+\frac{m_s\left[6\langle\bar{q}g_s\sigma
Gq\rangle-\langle\bar{s}g_s\sigma Gs\rangle \right]}{96\pi^2}\delta
(p_0-m_Q) +\frac{\langle\bar{q}q\rangle\langle\bar{s}s\rangle}{3}\delta(p_0-m_Q)\nonumber\\
&&+\frac{m_Q}{192\pi^2}\langle \frac{\alpha_sGG}{\pi}\rangle
\int_{t_i}^1 dt \left[ -3-2t+\frac{2}{t^2}\right]\nonumber \\
&& -\frac{m_Q}{384\pi^2}\langle \frac{\alpha_sGG}{\pi}\rangle
\int_0^1 dt\frac{(1-t)^2}{t}\widetilde{m}_Q\delta
(p_0-\widetilde{m}_Q) \, ,
\end{eqnarray}

\begin{eqnarray}
\rho^A_{\Sigma_Q}(p_0)&=&\frac{p_0}{32\pi^4}\int_{t_i}^1dt
t(1-t)^3(p_0^2-\widetilde{m}_Q^2)(5p_0^2-3\widetilde{m}_Q^2)+\frac{\langle\bar{q}q\rangle^2}{6}\delta(p_0-m_Q)\nonumber \\
&& +\frac{p_0}{96\pi^2}\langle \frac{\alpha_sGG}{\pi}\rangle
\int_{t_i}^1 dt (4-5t)\nonumber\\
&&+\frac{1}{192\pi^2}\langle \frac{\alpha_sGG}{\pi}\rangle
\int_{t_i}^1 dt (1-t)\widetilde{m}_Q^2\delta
(p_0-\widetilde{m}_Q)\nonumber \\
&& -\frac{m_Q^2}{288\pi^2}\langle \frac{\alpha_sGG}{\pi}\rangle
\int_0^1
dt\frac{(1-t)^3}{t^2}\left[1+\frac{\widetilde{m}_Q}{4T}\right]\delta
(p_0-\widetilde{m}_Q) \, ,
\end{eqnarray}

\begin{eqnarray}
\rho^B_{\Sigma_Q}(p_0)&=&\frac{3m_Q}{64\pi^4}\int_{t_i}^1dt
(1-t)^2(p_0^2-\widetilde{m}_Q^2)^2+\frac{\langle\bar{q}q\rangle^2}{3}\delta(p_0-m_Q)\nonumber\\
&&+\frac{m_Q}{192\pi^2}\langle \frac{\alpha_sGG}{\pi}\rangle
\int_{t_i}^1 dt \left[ -3-2t+\frac{2}{t^2}\right]\nonumber \\
&& -\frac{m_Q}{384\pi^2}\langle \frac{\alpha_sGG}{\pi}\rangle
\int_0^1 dt\frac{(1-t)^2}{t}\widetilde{m}_Q\delta
(p_0-\widetilde{m}_Q) \, ,
\end{eqnarray}
where $\widetilde{m}_Q^2=\frac{m_Q^2}{t}$,
$t_i=\frac{m_Q^2}{p_0^2}$.

\section*{Acknowledgements}
This  work is supported by National Natural Science Foundation,
Grant Number 10775051, and Program for New Century Excellent Talents
in University, Grant Number NCET-07-0282, and a foundation of NCEPU.

\end{document}